\documentclass[twocolumn,showpacs,preprintnumbers,amsmath,amssymb,pre]{revtex4}
\usepackage{graphicx}
\usepackage{setspace}
\usepackage{amsmath}
\usepackage{amssymb}
\usepackage{epsfig}
\usepackage{indentfirst}
\usepackage{bm}

\begin{document}

\title{Using dynamical barriers to control the transmission of light through \\
slowly-varying photonic crystals}

\author{A. J. Henning}
\affiliation{School of Physics and Astronomy, University of Nottingham, Nottingham NG7 2RD, UK}
\author{T. M. Fromhold}
\email{mark.fromhold@nottingham.ac.uk}
\affiliation{School of Physics and Astronomy, University of Nottingham, Nottingham NG7 2RD, UK}

\author{P. B. Wilkinson}
\affiliation{British Geological Survey, Kingsley Dunham Centre, Keyworth, Nottingham NG12 5GG, UK}

\date{\today}

\begin{abstract}
We use semiclassical Hamiltonian optics to investigate the propagation of light rays through two-dimensional photonic crystals when slow spatial modulation of the lattice parameters induces mixed stable-chaotic ray dynamics. This modulation changes both the shape and frequency range of the allowed frequency bands, thereby bending the resulting semiclassical ray trajectories and confining them within particular regions of the crystal. The curved boundaries of these regions, combined with the bending of the orbits themselves, creates a hierarchy of stable and unstable chaotic trajectories in phase space. For certain lattice parameters and electromagnetic wave frequencies, islands of stable orbits act as a dynamical barrier, which separates the chaotic trajectories into two distinct regions of the crystal, thereby preventing the rays propagating through the structure. We show that changing the frequency of the electromagnetic wave strongly affects the distribution of stable and unstable orbits in both real and phase space. This switches the dynamical barriers on and off and thus modulates the transmission of rays through the crystal. We propose microwave analogues of the photonic crystals as a route to the experimental study of the transport effects that we predict.  
\end{abstract}

\pacs{42.65.-k, 05.45.-a}

\maketitle

\section{Introduction}
Quantum chaos explores the quantum-mechanical properties of systems whose classical counterparts exhibit deterministic chaos \cite{Stoeckmann,TOM94}. Most studies of quantum chaos in experimentally-accessible systems have focused on atoms \cite{Stoeckmann,atom_billiards,37} and semiconductor systems, in which electron confinement by potential energy barriers of various shapes generates chaotic classical trajectories. 

Semiconductor superlattices, comprising a series of quantum wells, have been used to realize a fundamentally different regime of nonlinear dynamics, in which the effective classical Hamiltonian originates from an \emph{intrinsically quantum-mechanical} feature of the system: tunnel coupling of the wells to form energy bands, known as minibands \cite{FRO01,OSADA,FRO04,BAL2008,Green09}. When a bias voltage and tilted magnetic field are applied, the energy versus crystal momentum dispersion relation for the energy bands generates chaotic dynamics equivalent to those of a one-dimensional harmonic oscillator driven by a plane wave \cite{FRO01,FRO04,BAL2008,Green09}. Unusually, this system does not obey the KAM theorem \cite{ZAS91,ZAS2004,Luo04,Karney,Soskin09}, which means that chaos switches on and off abruptly at critical values of the applied fields. At the critical fields, the electron phase space is threaded by intricate web patterns, known as stochastic webs \cite{ZAS91,ZAS2004,Luo04,Karney,Soskin09}. These webs form a network of conduction channels, through which the electrons propagate in real space \cite{FRO01,OSADA,FRO04,BAL2008,Green09}. When the web is switched on, the electrons undergo chaotic diffusive motion along its filaments, thereby producing a sharp increase in the measured and calculated direct current \cite{FRO04,BAL2008} and increasing both the power and frequency of GHz/THz current oscillations by an order of magnitude \cite{Green09}. Consequently, chaotic semiclassical transport through energy bands provides a mechanism for controlling transport through periodic structures. In the absence of electron scattering, switching due to the formation and destruction of stochastic webs is extremely sensitive, due to the inherent instability of chaotic orbits, and would produce $\delta-$function peaks in the current-voltage characteristics. In real superlattices, though, the peaks are broadened by electron scattering. This limitation can, in principle, be overcome by using an analogous system comprising ultracold atoms moving through the energy bands of an optical lattice with a tilted harmonic trap \cite{Scott:02}. 

Due to the formal similarity between Maxwell's equations and the Schr\"odinger equation \cite{Stoeckmann}, microwave and optical systems have been used to study the ray-wave correspondence, analogous to quantum chaos, for electromagnetic radiation. Much of this work has focused on electromagnetic billiards in which the radiation is confined in a quasi two-dimensional (2D) region of space, usually a cavity or dielectric block, whose boundary is shaped to create chaotic dynamics for light rays confined within the block. There has been considerable technological interest in such systems because, when used as a resonator for a micro laser, they can produce strongly focused emission whose power is orders of magnitude higher than obtained from traditional circular resonators \cite{Gmachl_1998, Lee_2004, Mekis_1995, Schwefel_2004, Nockel_1994}. Microwave systems have also been used to study the ray-wave correspondence in billiards, photonic crystals, and in a range of disordered dielectric media \cite{Stein_1995, Stockmann_1990, Kuhl_2000,StockNJP}. 

In 2D photonic crystals \cite{shep1,PhysRevA.46.612}, ray dynamics related to the chaotic semiclassical motion of electrons in a superlattice miniband has been predicted when the lattice parameters, and hence the local refractive index, vary slowly with position \cite{crd, CTS}. Such structures offer major advantages over superlattices for studying, and potentially exploiting, chaotic band transport effects because they operate at room temperature and without an applied magnetic field. As a ray passes through the crystal, changes in the lattice parameters alter the local photonic band structure, in particular the shape and frequency range of the allowed bands. This exerts an effective \textquotedblleft force" on the ray \cite{pboawsl, crd}, whose trajectory can be determined by solving Hamilton's equations \cite{crd}, just like calculating the path of a classical particle subject to real forces. When the lattice parameters are chosen to approximate the potential landscape experienced by an electron in a superlattice with a tilted magnetic field, the ray paths share many characteristics of non-KAM chaos. In particular, they have a rich mixed stable-chaotic phase space structure \cite{crd}. 

\begin{figure}[t]
\centerline{
   \includegraphics[width =\linewidth]{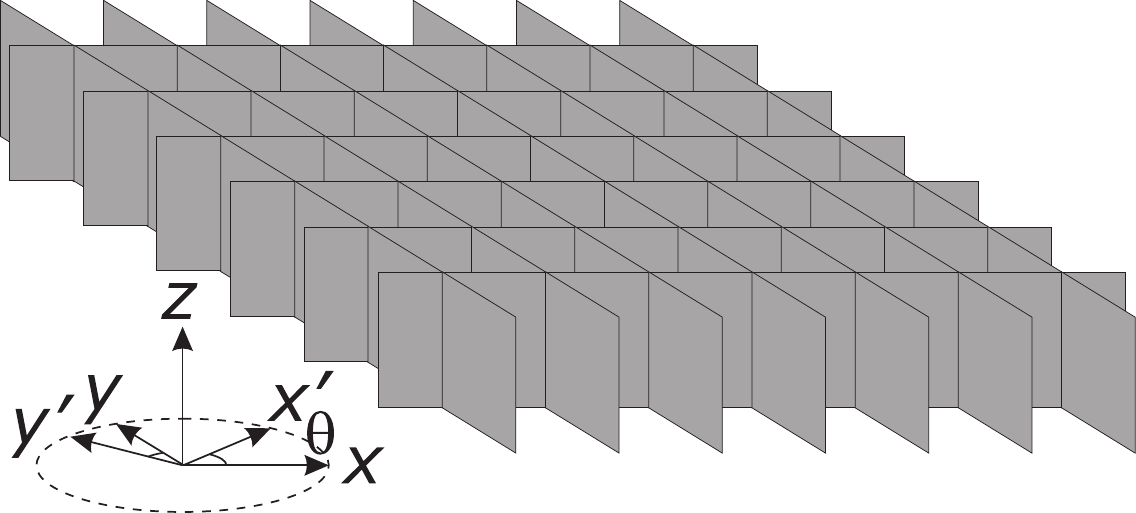}
  }
	\caption{Schematic diagram of the photonic crystal, which comprises two arrays of parallel dielectric sheets (gray)  intersecting at 90$^{\circ}$. Each sheet is of infinite area. The sheets are parallel to the $x$-$z$ or $y$-$z$ plane. The $x'$ and $y'$ axes are related to the $x$ and $y$ axes by a rotation of $\theta$ about the $z$ axis (inset).}
	\label{fig1}
\end{figure}

In this paper, we study the origin of mixed-stable-chaotic ray dynamics in spatially-modulated photonic crystals and how such dynamics depend on the parameters both of the crystal and the electromagnetic wave. We focus on two key aspects of the ray dynamics. Firstly, we consider how the spatial variation of the lattice parameters affects the range and dispersion of the allowed frequency bands and, hence, the ray trajectories obtained from Hamiltonian ray dynamics. Secondly, we analyze how the form of the ray trajectories, in real and phase space, depends on the frequency, $\omega$, of the electromagnetic wave. In contrast to previous work \cite{crd}, which considered ray dynamics at fixed frequency, we find that the location and extent of the stable and chaotic orbits depend strongly on $\omega$. By considering four critical points in the local Brillouin zone, we determine the loci of classically-allowed regions of the photonic crystal, where rays can propagate, and the corresponding forbidden regions that the rays cannot enter. Within the classically allowed regions, we identify additional \emph{dynamical} barriers, which are formed by localized stable orbits and are impenetrable to chaotic ray trajectories \cite{crd}. We show that as the ray propagates through the crystal, changes in the local bandstructure cause it to reach the band extrema (highest or lowest frequency within an allowed band), where Bragg reflection occurs. Together, Bragg reflection and the action of the dynamical barriers limit the spatial region in which a ray moves, and drive many ray paths chaotic. As $\omega$ increases, a single region of chaos splits into two distinct chaotic seas separated by a dynamical barrier. Further increasing $\omega$ suppresses chaos in one of these two regions, thereby deactivating the dynamical barrier and its effect on ray transport.

The structure of the paper is as follows. In Section \ref{sec:The structure of the crystal}, we define the crystal structure and consider the frequency-versus Bloch wavevector dispersion relations for the allowed frequency bands in which electromagnetic radiation can propagate through the crystal. Next, in Section \ref{sec:The ray tracing} we introduce the semiclassical (Hamiltonian) equations of ray motion and consider, qualitatively, how the shape of the dispersion relations influences the ray trajectories, in particular whether they are open or closed. In Section \ref{sec:The shape of the ray paths}, we consider the spatial regions of the crystal in which the rays propagate and show how the shapes of these regions, and the orbits themselves, are determined by the form of the dispersion relations and the photonic crystal structure. In Section \ref{sec:Dynamical Barriers}, we demonstrate the existence and influence of dynamical barriers, which originate from islands of stability within a chaotic sea rather than from a classically forbidden region of high potential energy. In particular, we show that these dynamical barrriers can control the transmission of electromagnetic waves through the photonic crystal. In Section \ref{sec:experiment}, we discuss possible routes to the experimental realization and study of the dynamical regimes that we introduce. Finally, in Section \ref{sec:Summary}, we summarize our results and draw conclusions.

\section{The structure of the crystal}
\label{sec:The structure of the crystal}
We consider photonic crystals created from two intersecting sets of dielectric sheets (see Fig. \ref{fig1}). One set is parallel to the $x$-$z$ plane and the other is parallel to the $y$-$z$ plane. The crystal is invariant, and taken to be infinite, along the $z$ axis. This creates a photonic crystal in which the dielectric sheets enclose rectangular air gaps, as shown in Fig. \ref{fig1}.

As in Refs. [\onlinecite{shep1,PhysRevA.46.612}], we characterize each dielectric sheet by the dielectric parameter $m=(\epsilon_{r}-1)d$, where $d$ and $\epsilon_{r}$ are, respectively, the width and relative permittivity of the sheet. At the lines where two sheets intersect, the relative permittivity is $2\epsilon_{r}-1$. At position $\mathbf{r}=(x, y)$ within the crystal, we take the local unit cell length in the $x$ and $y$ directions to be $l_{x}$ and $l_{y}$ respectively. In the limit $d$$\rightarrow$0, if the electric field vector of an electromagnetic wave of frequency $\omega$ is parallel to the $z$ axis, the local frequency versus wavevector dispersion relation, $\omega_{\mathrm{loc}}(\bm{\mu},l_{x},l_{y})$, of the wave satisfies the equations \cite{shep1}, 

\begin{subequations}
\begin{align}
\label{eqna}
\cos\left(\mu_{x}l_{x}\right)=\cos\left(k_{x}l_{x}\right)-\frac{m\omega_{\mathrm{loc}}^{2}}{2c^{2}k_{x}}\sin\left(k_{x}l_{x}\right)\\
\cos\left(\mu_{y}l_{y}\right)=\cos\left(k_{y}l_{y}\right)-\frac{m\omega_{\mathrm{loc}}^{2}}{2c^{2}k_{y}}\sin\left(k_{y}l_{y}\right),
\label{eqnb}
\end{align}
\end{subequations}
where $c$ is the speed of light in vacuo, $\bm{\mu}=$($\mu_{x}$,$\mu_{y}$) is the Bloch wavevector, and $\bm{k}$=($k_{x}$,$k_{y}$) is the local wavevector of the electromagnetic wave between the dielectric sheets. At time $t$, the local electric field magnitude, $E = E_{0}e^{i(\mathbf{k}.\mathbf{r}-\omega{t})}$, is determined by the wavevector $\mathbf{k}$. The Bloch wavevector, $\bm{\mu}$, determines the spatial development of the phase of the electromagnetic wave as it propagates through the crystal \cite{shep1}. 

When the $y$-component of the wavevector between the dielectric sheets is evanescent, so that $k_{y} = iq_{y}$ where $q_{y}$ is real, Eq. (\ref{eqnb}) can be written in the form \cite{shep1}:

\begin{equation}	\cos\left({\mu}_{y}l_{y}\right)=\cosh\left({q}_{y}l_{y}\right)-\frac{m\omega_{\mathrm{loc}}^{2}}{2c^{2}{q}_{y}}\sinh\left({q}_{y}l_{y}\right).
\label{evenes}
\end{equation}

If the lattice parameters $m$, $l_{x}$ and $l_{y}$ vary sufficiently slowly, the local dispersion relation at any point in the crystal is, to good approximation, the same as the dispersion relation for an infinite crystal with a constant unit cell identical to the local unit cell \cite{crd}. Here, we choose $m$ and $l_{x}$ values that vary slowly with position and are defined by the continuous functions $l_{x} = l_{0}\exp(-\eta{x})$ and $m=m_{0}(1-\rho{y'}^{2})$, where the constants $l_{0}$, $m_{0}$, $\eta$, and $\rho$ are specified below. The $x' - y'$ axes, are related to the $x - y$ axes by a rotation of $\theta$ about the $z$ axis, as shown in Fig. \ref{fig1}.

The solutions of Eqs.(\ref{eqna}) and (\ref{evenes}) define a number of allowed frequency bands of the crystal. In the bands considered here, we show below that changes in the positions of both the top and bottom of the band, i.e. the highest and lowest frequencies, play a key role in driving the ray paths chaotic. For that reason, the lowest frequency band is of no interest here, because it always has a solution extending down to $\omega =0$, which means that chaotic ray paths within the band are unbounded on the right (i.e. for high $x$) \cite{crd}. Instead, the band of interest in this work is the next lowest frequency band in which the rays have a locally evanescent component in the $y$ direction and a locally propagating character in the $x$ direction \cite{crd,shep1}. 

\section{The ray tracing}
\label{sec:The ray tracing}

The ray paths in the crystal are determined by the pair of Hamilton's equations \cite{crd,PhysRevE.70.036612}
\begin{subequations}
\begin{equation}
\frac{d\mathbf{r}}{dt}=\frac{\partial{H}}{\partial\bm{\mu}},\label{hamil1}
\end{equation}
\begin{equation}
{\frac{d\bm{\mu}}{dt}=-\frac{\partial{H}}{\partial\mathbf{r}}}.
	\label{hamil}
\end{equation}
\end{subequations}

\begin{figure}[t]
\centerline{
   \includegraphics[width =\linewidth]{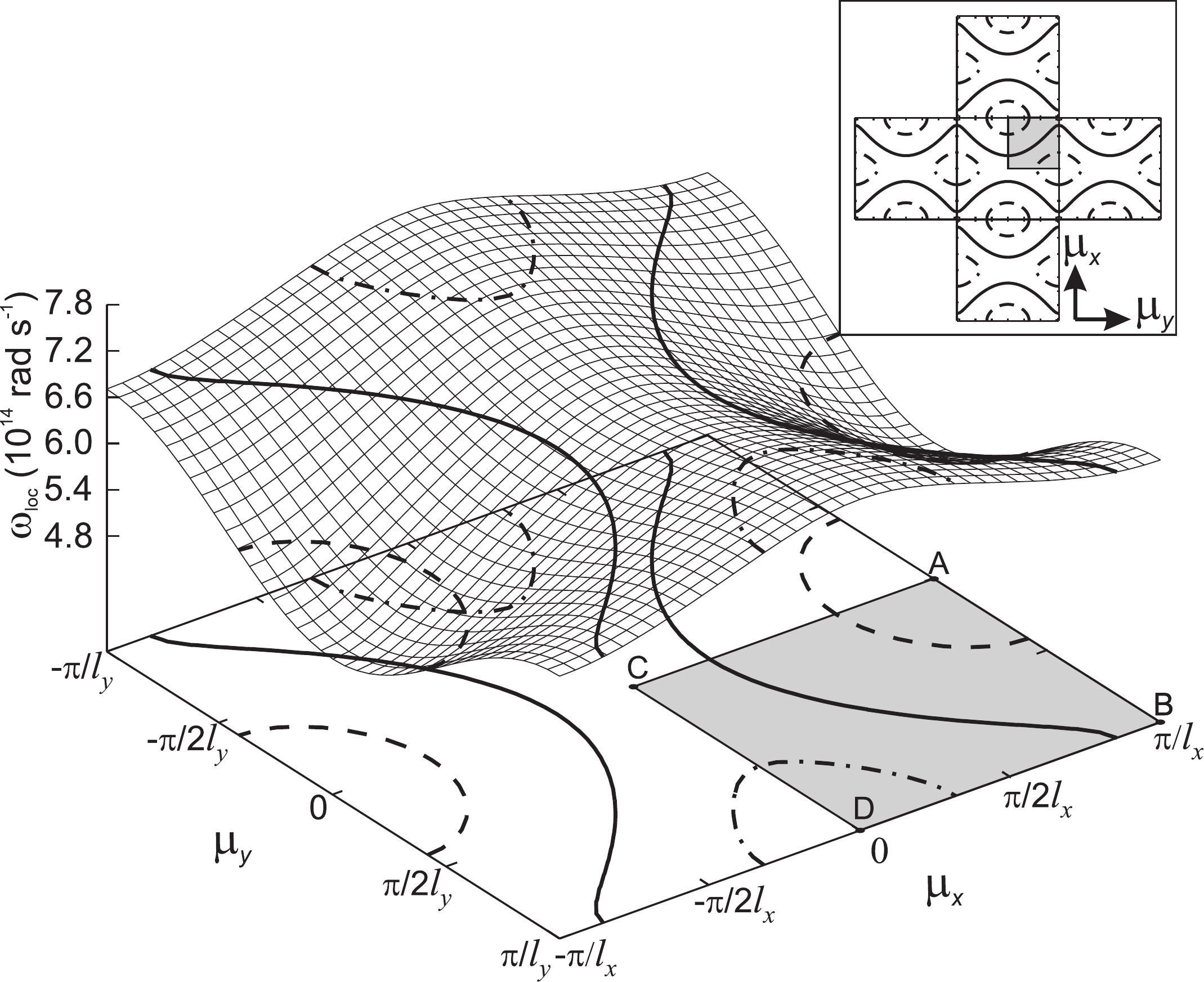}
  }
	\caption{Dispersion surface, $\omega_{\mathrm{loc}}(\mu_{x},\mu_{y})$ calculated for $l_{x}=0.75$ $\mu$m, $l_{y}=0.5$ $\mu$m and $m_{0}=2$ $\mu$m. Three constant frequency contours are shown with $\omega = 5.7\times10^{14}$ rad s$^{-1}$ (dashed curve), $\omega = 6.75\times10^{14}$ rad s$^{-1}$ (solid curve) and $\omega = 7.35\times10^{14}$ rad s$^{-1}$ (dot-dashed curve). At the bottom of the figure the contours are shown projected onto the $\omega_{\mathrm{loc}}=0$ plane. Rays propagate perpendicular to these contours. One quadrant of the reduced Brillouin zone is shaded gray with its corners labelled A, B, C, D. The inset, top right, shows the repetition of the Brillouin zone and illustrates that the lowest and highest frequency contours (dashed and dot-dashed curves respectively) form closed curves while the intermediate contour (solid) forms an open curve.}
	\label{fig2}
\end{figure}

The 2D Hamiltonian is defined as \cite{crd}
\begin{equation}
	H(\mathbf{r},\bm{\mu},\omega)=\omega_{\mathrm{loc}}(\bm{\mu},l_{x},l_{y})-\omega,
	\label{disp}
\end{equation}where $\omega$ is the angular frequency of the electromagnetic wave and the local lattice constants $l_{x},l_{y}$ both depend on position \textbf{r}. To determine ray paths in the second band we first specify $\omega$ and the initial position of the ray front, both spatially in the crystal, ($x,y$), and the Bloch wavevector in the band, ($\mu_{x},\mu_{y}$). We then solve Eqs. (\ref{hamil1}) and (\ref{hamil}) numerically using a fourth-order Runge-Kutta method. It follows from Eqs.(\ref{hamil1}) and (\ref{disp}) that
\begin{equation}
	\frac{d\mathbf{r}}{dt}=\frac{\partial{\omega_{\mathrm{loc}}}}{\partial\bm{\mu}}.
	\label{pathexp}
\end{equation}

From Eq.(\ref{pathexp}) we see that the direction in which a ray propagates will always be perpendicular to the constant frequency contour $\omega = \omega_{\mathrm{loc}}$ on the dispersion surface. 

Figure \ref{fig2} shows $\omega_\mathrm{loc}$ versus $\mu_{x}$ and $\mu_{y}$ at the point ($x,y$) = ($28.7$ $\mu$m, $y=11$ $\mu$m), in the crystal where $l_{x}=0.75$ $\mu$m, $l_{y}=0.5$ $\mu$m, and $m_{0}=2 $ $\mu$m. We also take $\theta = 21^{\circ}$ and $\eta$ = $10^{4}$ m$^{-1}$: parameters that will be used throughout this paper. The three constant frequency contours marked in 
Fig. \ref{fig2} correspond to $\omega = 5.7\times10^{14}$ rad s$^{-1}$ (dashed curve), $\omega = 6.75\times10^{14}$ rad s$^{-1}$ (solid curve) and $\omega = 7.35\times10^{14}$ rad s$^{-1}$ (dot-dashed curve). 

Due to the symmetry of the frequency band, we need only consider one quadrant of the reduced Brillouin zone (gray shaded region in Fig. \ref{fig2}) for which $0\leq\mu_{x}\leq\pi/l_{x}$ and $0\leq\mu_{y}\leq\pi/l_{y}$. In fact, it is the four corners (marked A, B, C and D in Fig. \ref{fig2}) of this reduced Brillouin zone quadrant i.e. the points where $\mu_{x}=0$ or $\pi/l_{x}$ and $\mu_{y}=0$ or $\pi/l_{y}$ that will be of most interest here.

The inset in Fig. \ref{fig2} reveals that, since the Brillouin zone repeats with a period of $2\pi/l_{x}$ in the $\mu_{x}$ direction and $2\pi/l_{y}$ in the $\mu_{y}$ direction, the lowest frequency contour at $\omega = 6.75\times10^{14}$ rad s$^{-1}$ (dashed curve), and the highest frequency contour at $\omega = 7.35\times10^{14}$ rad s$^{-1}$ (dot-dashed curve), both form closed curves. However, this is not the case for the intermediate contour at $\omega = 6.75\times10^{14}$ rad s$^{-1}$ (solid curve), which is open in the extended zone scheme. 

Since the frequency band varies monotonically in both the $\mu_x$ and $\mu_y$ directions over the shaded section of the reduced Brillouin zone in Fig. \ref{fig2}, the gradient $\partial\omega_\mathrm{loc}/\partial\bm{\mu}$ is only parallel to the $\mu_{x}$ axis along the lines $\mu_{y}=0$ or $\pi/l_{y}$, and only parallel with the $\mu_{y}$ axis along the lines $\mu_{x}=0$ or $\pi/l_{x}$. When contours form open curves they are restricted to lie either between $\mu_{y}=0$ and $\pi/l_{y}$, or between $\mu_{x}=0$ and $\pi/l_{x}$. Since a ray propagates perpendicular to the constant frequency contour where $\omega_\mathrm{loc} = \omega$, if this contour forms an open curve there are directions in which the ray cannot propagate. For example, it can be seen from Fig. \ref{fig2} that for the (solid) constant frequency contour at $\omega = 6.75\times10^{14}$ rad s$^{-1}$, rays cannot propagate parallel, or almost parallel, to the $y$ axis, because $\omega_\mathrm{loc}$ is never parallel to the $\mu_{x}$ axis. Below, we show that this restriction causes spatial localization of the rays in the $y$ direction.

For a given $\omega$, a constant frequency contour forms a closed loop if $\omega = \omega_\mathrm{loc}$ is between the bottom of the band, point A in Fig. \ref{fig2}, and the saddle point B. The contour is also closed if $\omega_\mathrm{loc}$ is between the top of the band, point D in Fig. \ref{fig2}, and the saddle point C. The constant frequency contours between the two frequencies corresponding to $\omega_\mathrm{loc}$ at points B and C always forms open curves.

\section{The shape of the ray paths}
\label{sec:The shape of the ray paths}
Henceforth, we consider crystal parameters, $l_{y}= 1$ $\mu$m, $l_{0} =1$ $\mu$m, $\theta = 21^{\circ}$, $\eta=1\times10^{4}$ m$^{-1}$, $m_{0}=2$ $\mu$m, $\rho =1\times10^{10}$ m and, unless otherwise stated, take $\omega =$ $6\times10^{14}$ rad s$^{-1}$. When there is an evanescent component to the electric field, this is in the $y$ direction. 

Spatial variation of the unit cell changes both the shape of the dispersion curve, $\omega_{\mathrm{loc}}(\bm{\mu},l_{x},l_{y})$, and the range of frequencies that it spans. Thus, in parts of the crystal the electromagnetic wave may lie within an allowed band,  but in other parts its frequency coincides with a bandgap.

It is therefore essential to identify where in the crystal $\omega$ coincides with an extremum (points A and D in Fig. \ref{fig2}) or a saddle point (points B and C in Fig. \ref{fig2}) in the dispersion surface.

\begin{figure}[t]
\centerline{
   \includegraphics[width =\linewidth]{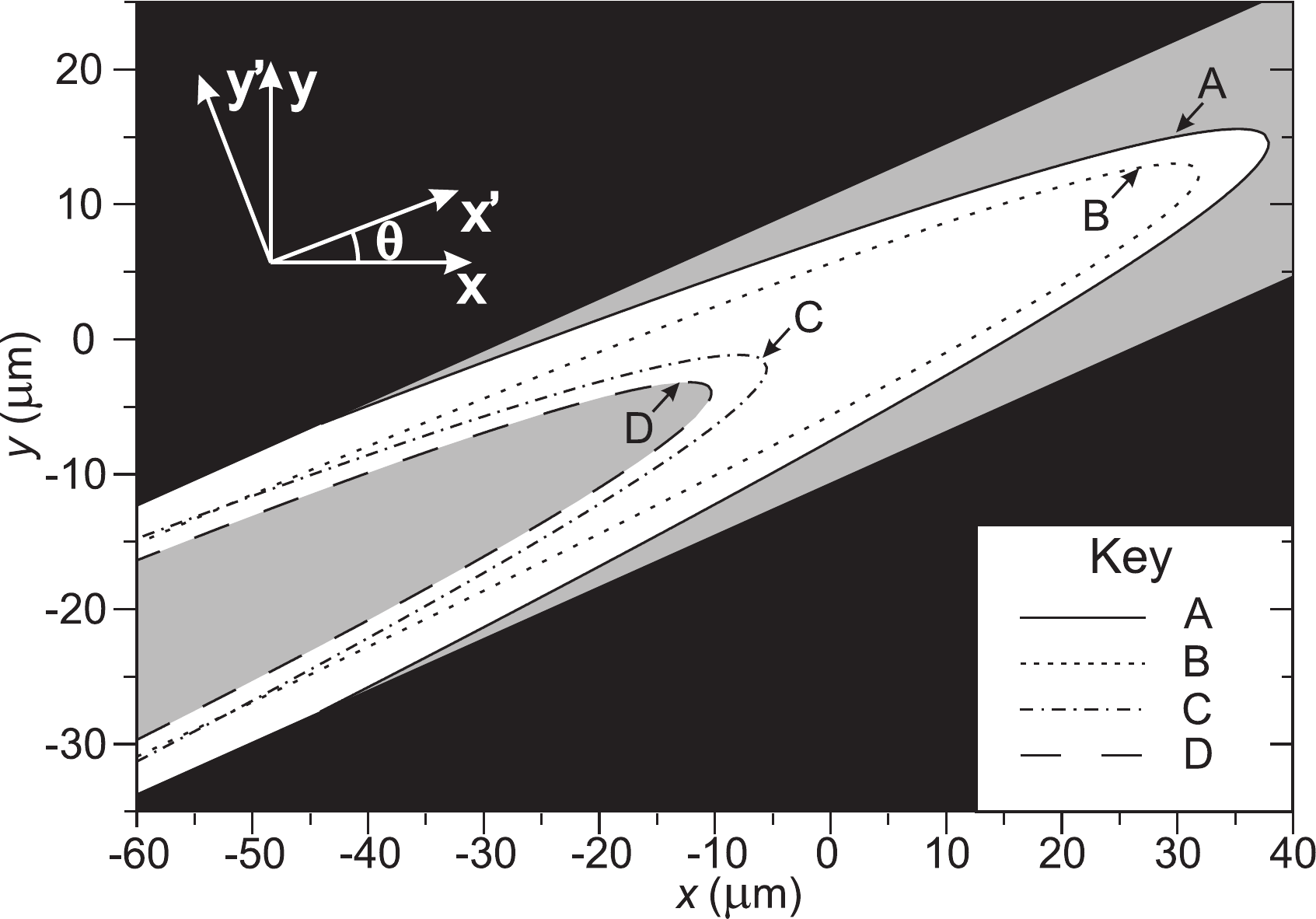}
  }
	\caption{Loci of points in the crystal where $\omega=$ $6\times10^{14}$ rad s$^{-1}$ coincides with the $\omega_{\mathrm{loc}}$ values at points A (solid curve), B (dotted curve), C (dot-dash curve) and D (dashed curve) in the reduced Brillouin zone shown in Fig. \ref{fig2}. In the gray regions, $\omega$ lies in a bandgap, and hence there are no propagating solutions. Rays can only propagate in the white regions bounded by the solid and dashed curves where $\omega$ coincides with the band minimum and maximum, respectively, shown in Fig. \ref{fig2}. Within the black regions, $m$ is negative, meaning that either $\epsilon_{r}$ or $d$ must also be negative. Rays cannot reach these regions of negative permittivity because they are unable to traverse the outer gray area where $\omega$ lies within a bandgap. Consequently, the regions of negative permittivity do not affect the ray dynamics that we consider.}
	\label{fig3}
\end{figure}

To illustrate this, Fig. \ref{fig3} shows the four loci of points in the crystal at which $\omega$ coincides with the two extrema or the two saddle points in the local $\omega_{\mathrm{loc}}(\bm{\mu},l_{x},l_{y})$ surface (i.e. points A-D in Fig. \ref{fig2}). The solid, dotted, dot-dashed and dashed curves correspond to points A, B, C and D respectively. Between curves A and D, the white region in Fig. \ref{fig3}, $\omega$ lies within the second allowed frequency band. Whilst it is not universally true that point B lies at a lower frequency than point C, as can be seen from the crossing of curves B and C at $x = x_{cross} \approx -48$ $\mu$m in Fig. \ref{fig3}, this is the case throughout the region of interest here. The gray region in Fig. \ref{fig3} shows the regions of the crystal for which there are no propagating solutions at frequency $\omega$. In the black region, $m<0$, meaning that either $\epsilon_{r}$ or $d$ would need to be negative. Although some materials do have negative values of permittivity, we do not consider them here.  

As discussed above, at any point in the crystal a ray can, in principle, propagate in any spatial direction provided that the $\omega_{\mathrm{loc}} = \omega$ contour forms a closed loop. By contrast, the range of possible propagation directions is limited for open contours. In Fig. \ref{fig3}, when $x > x_{cross}$, at any point in the crystal that lies between curves A and B, or between curves C and D, the iso-frequency contour for $\omega_\mathrm{loc} = \omega =$ $6\times10^{14}$ rad s$^{-1}$ forms a closed loop. Conversely, for any point in the crystal between curves B and C, the iso-frequency contour forms an open curve, which restricts the range of directions in which a ray may propagate.  
\begin{figure}[t]
\centerline{
   \includegraphics[width =0.9\linewidth]{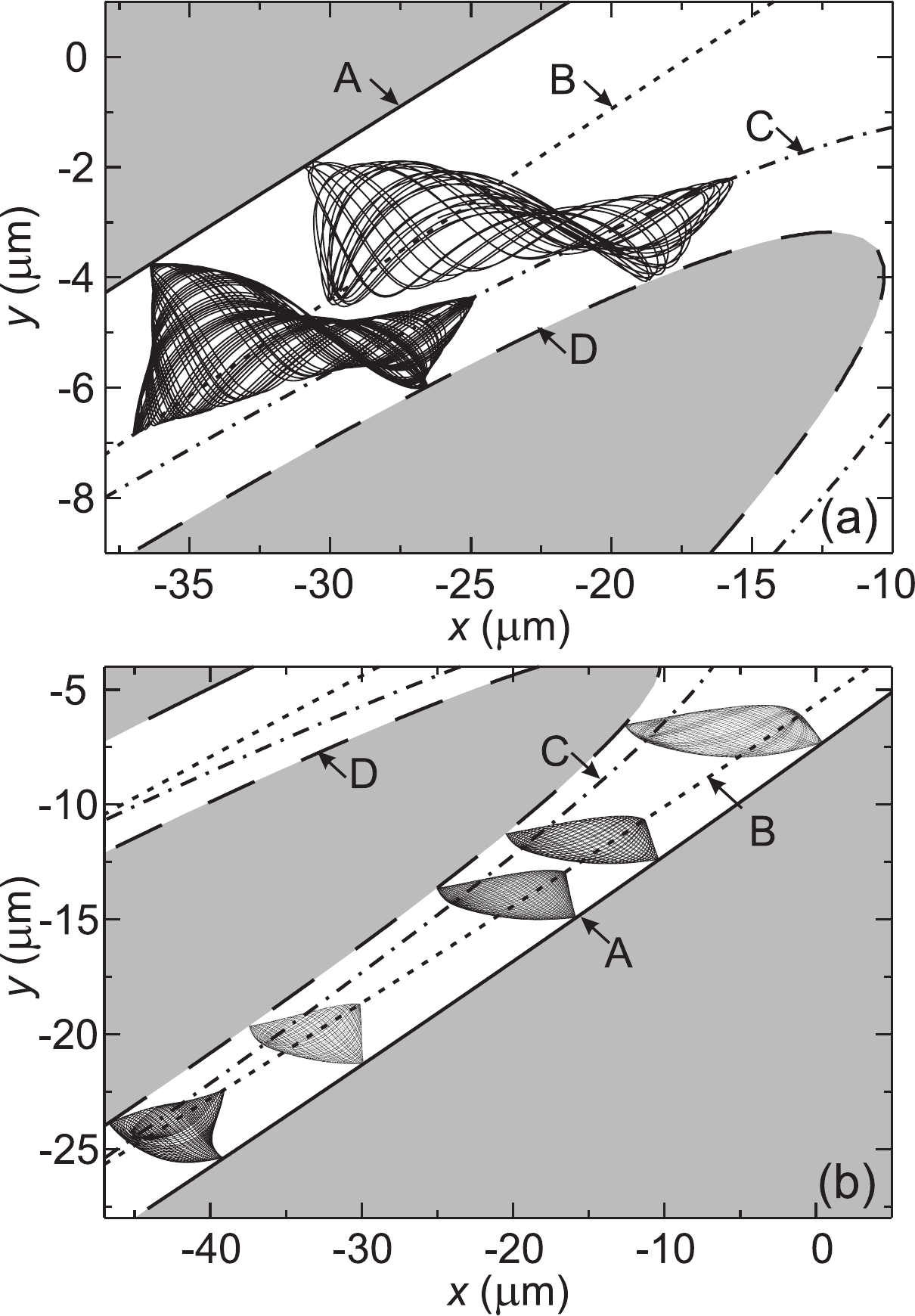}
  }
	\caption{Ray paths calculated in a region of the crystal where all trajectories are stable. The four curves are the loci of points in the crystal where $\omega=$ $6\times10^{14}$ rad s$^{-1}$ coincides with the $\omega_{\mathrm{loc}}$ values at points A (solid curve), B (dotted curve), C (dot-dash curve) and D (dashed curve) in the reduced Brillouin zone. These four curves strongly influence the shape of the ray paths and determine the turning points.}
	\label{fig4}
\end{figure}

\begin{figure}[t]
\centerline{
  \includegraphics[width =\linewidth]{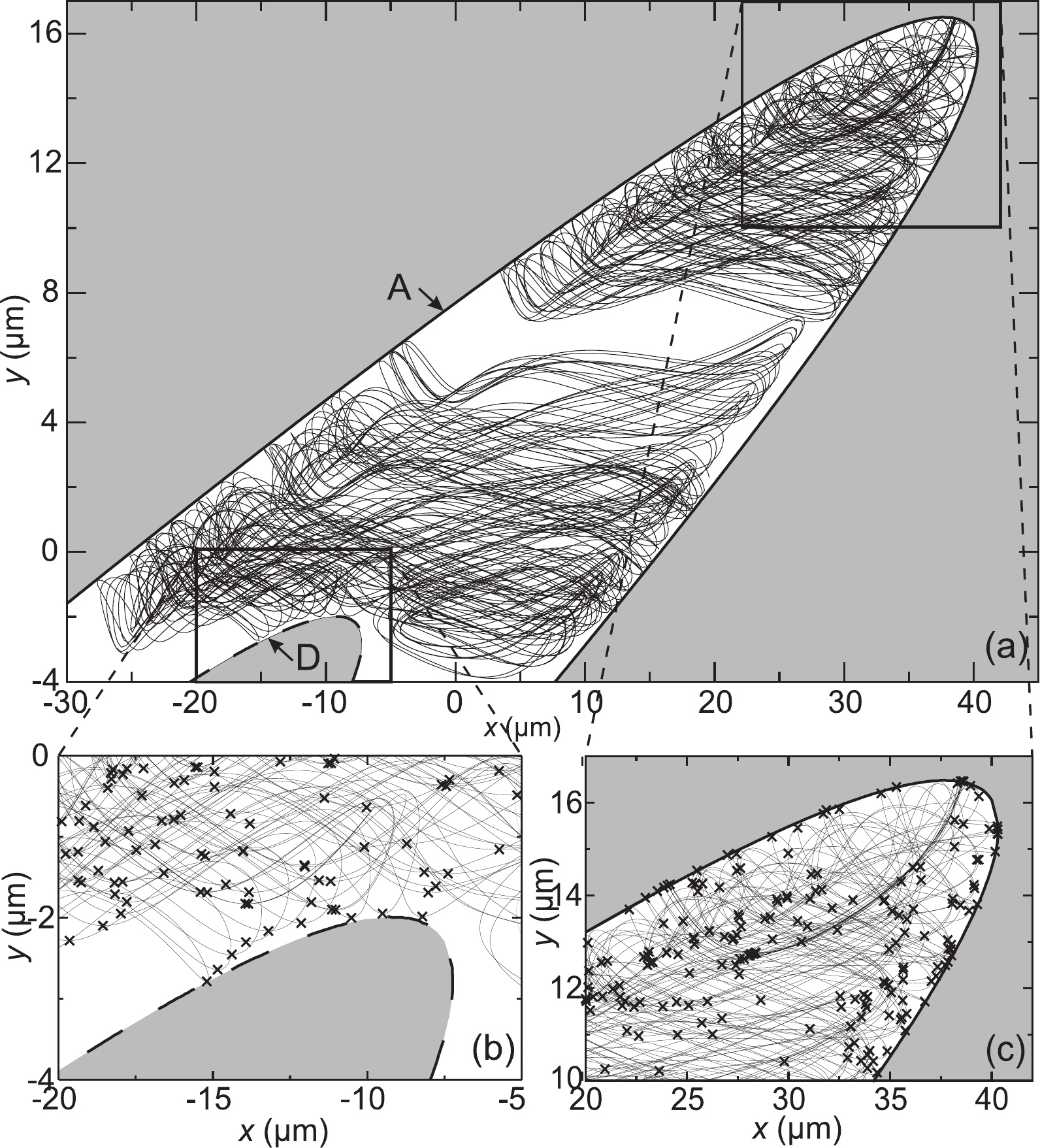}
  }	
\caption{(a) Two chaotic ray paths propagating in two distinct regions of the photonic crystal when $\omega = $ $6.09\times10^{14}$ rad s$^{-1}$. The crystal parameters are $l_{y}= 1$ $\mu$m, $l_{0} =1$ $\mu$m, $\theta = 21^{\circ}$, $\eta=1\times10^{4}$ m$^{-1}$, $m_{0}=2$ $\mu$m, $\rho =1\times10^{10}$ m. Panels (b) and (c) are enlargements of the left and right-hand boxes in (a). Crosses in (b) and (c) mark the points where Bragg reflections occur.}
	\label{fig5}
\end{figure}

Figure \ref{fig4} shows a selection of stable ray paths in the crystal. The turning points of these rays are produced either by Bragg reflection, whenever $\mu_{x}$ = $\pm\pi/l_{x}$ or $\mu_{y}$ = $\pm\pi/l_{y}$, or by the rays reaching band extrema away from the Brillouin zone boundaries as a result of the band changing shape throughout the crystal. Figure \ref{fig4} reveals that the ray paths are shaped by the shifting of the bandstructure, with curves A, B, C and D passing through extrema in the envelopes of the ray trajectories. Between curves A and B, Bragg reflections occur when $\mu_{x}=\pm\pi/l_{x}$. At frequencies above that at point B in Fig. \ref{fig2}, such reflections no longer occur because the iso-frequency contours in Fig. \ref{fig2} only intersect with the edge of the Brillouin zone where $\mu_{y}=\pm\pi/l_{y}$ (between points B and D).

\section{Dynamical Barriers}
\label{sec:Dynamical Barriers}
Previous work has shown that the phase space of rays in the photonic crystal contains mixed stable-chaotic regions and that, for certain $\omega$, two or more chaotic regions are separated by islands of stability \cite{crd}. A trajectory starting in one chaotic sea cannot cross the stable island and is therefore unable to enter the second chaotic sea. Consequently, the stable island is known as a dynamical barrier. Although there is sometimes diffusion across dynamical barriers, such as via cantori (see, for example, [\onlinecite{TTC}]), in the photonic crystals considered here, the dynamical barrier can be completely impenetrable, localizing chaotic rays to two distinct regions of the phase space.

In this section, we explain the existence of these dynamical barriers and explore their effect on ray transport. The dynamical barriers occur for a wide range of $\omega$ and, at the higher frequencies, separate the chaotic regions in real space as well as in phase space. Consequently, dynamical barriers provide a mechanism for controlling the transmission of light through photonic crystals.

Figure \ref{fig5}(a) shows two distinct chaotic paths in the photonic crystal that are almost separated in real space.  Figures \ref{fig5}(b) and (c) show enlargements of the regions within the boxes in (a), in which the rays are driven chaotic. The solid curve, (A), in Fig. \ref{fig5} marks the loci of points where the bottom of the frequency band (point A in Fig. \ref{fig2}) is at $\omega_\mathrm{loc}=$ $6.09\times10^{14}$ rad s$^{-1}$, whilst the dashed curve, (D), marks the loci of points where the \emph{top} of the band (point D in Fig. \ref{fig2}) is at $\omega_{\mathrm{loc}}=$ $6.09\times10^{14}$ rad s$^{-1}$. 

The crosses in Fig. \ref{fig5} (b,c) mark the points along the real-space ray trajectory where the Bloch wavevector, $\bm{\mu}$, is at an edge of the reduced Brillouin zone shown in Fig. \ref{fig2}. At such points, the electromagnetic wave Bragg reflects along either the $x$ or $y$ direction, corresponding to a turning point of the classical ray. Most of these turning points occur away from the frequency band extrema labelled A, B and D in Fig. \ref{fig2} and therefore also away from the edges of the classically-allowed (white) region in Fig. \ref{fig5} (b,c). But, occasionally, Bragg reflection occurs at, or very near to, one of these extrema. For example, the crosses situated exactly on the dashed curve in Fig. \ref{fig5} (b), which is the locus of points in real space along which the electromagnetic wave frequency $\omega$ coincides with the top of the frequency band at Point D in Fig. \ref{fig2}, correspond to Bragg reflections along the $y$ direction occurring at Point D. 

\begin{figure}[t]
\centerline{
   \includegraphics[width =0.95\linewidth]{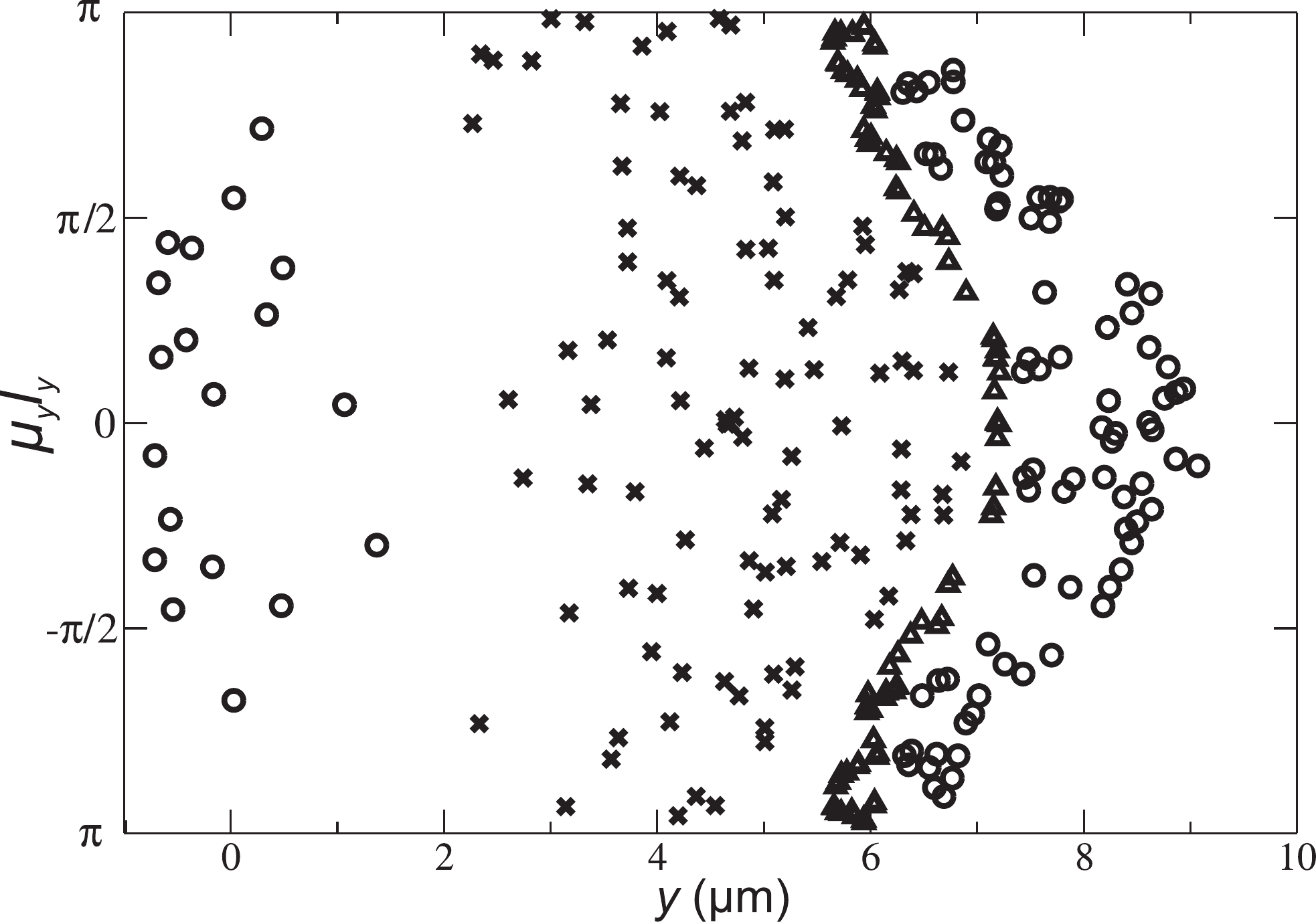}
  }
	\caption{A Poincar\'e section in which a point is plotted each time a ray turns from traveling along the negative to the positive $x$-direction. The points are generated by three ray paths: the two chaotic trajectories shown in Fig. \ref{fig5} and a stable ray path that separates them. Points produced by the upper path in Fig. \ref{fig5} are marked with crosses, while those produced by the lower path are marked with circles. Points generated by a stable path in the dynamical barrier are marked with triangles.}
	\label{fig6}
\end{figure}

\begin{figure*}[t]
\centerline{
   \includegraphics[width =\linewidth]{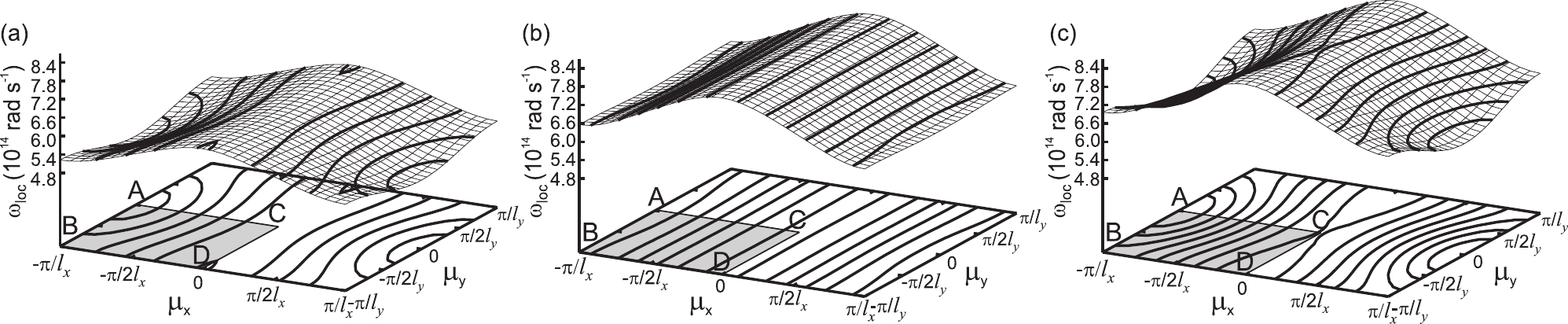}
  }
	\caption{Dispersion surfaces $\omega_\mathrm{loc}(\mu_{x},\mu_{y})$ calculated at three different positions in the photonic crystal: (a) $x=0$ $\mu$m, $y'=0$ $\mu$m where $l_{x} = l_{y} = 1$ $\mu$m and $m=2$ $\mu$m; (b) $x=50$ $\mu$m, $y'=0$ $\mu$m where $l_{x} = 0.607$ $l_{0}$ $\mu$m, $l_{y} = 1$ $\mu$m and $m=2$ $\mu$m; (c) $x=0$ $\mu$m, $y'=7.07$ $\mu$m where $l_{x} = l_{y} = 1$ $\mu$m and $m=1$ $\mu$m. The iso-frequency contours, shown both on the dispersion surface and projected onto the $\omega_\mathrm{loc}=0$ plane beneath, are equally spaced by $3\times10^{13}$ rad s$^{-1}$. As $l_{x}$ decreases, the band shifts to higher $\omega_\mathrm{loc}$, as can be seen by comparing the dispersion surface in (b) with that in (a). In addition, over large sections of the dispersion surface in (b), the iso-frequency countours are almost parallel to the $\mu_{y}$ axis, meaning that rays are, in general, restricted to propagate in directions close to the $x$ axis. Comparing (c) to (a) reveals that as $m$ decreases, the band shifts to higher $\omega_\mathrm{loc}$ and also broadens.}
	\label{fig7}
\end{figure*}

\begin{figure}[b]
\centerline{
   \includegraphics[width =0.7\linewidth]{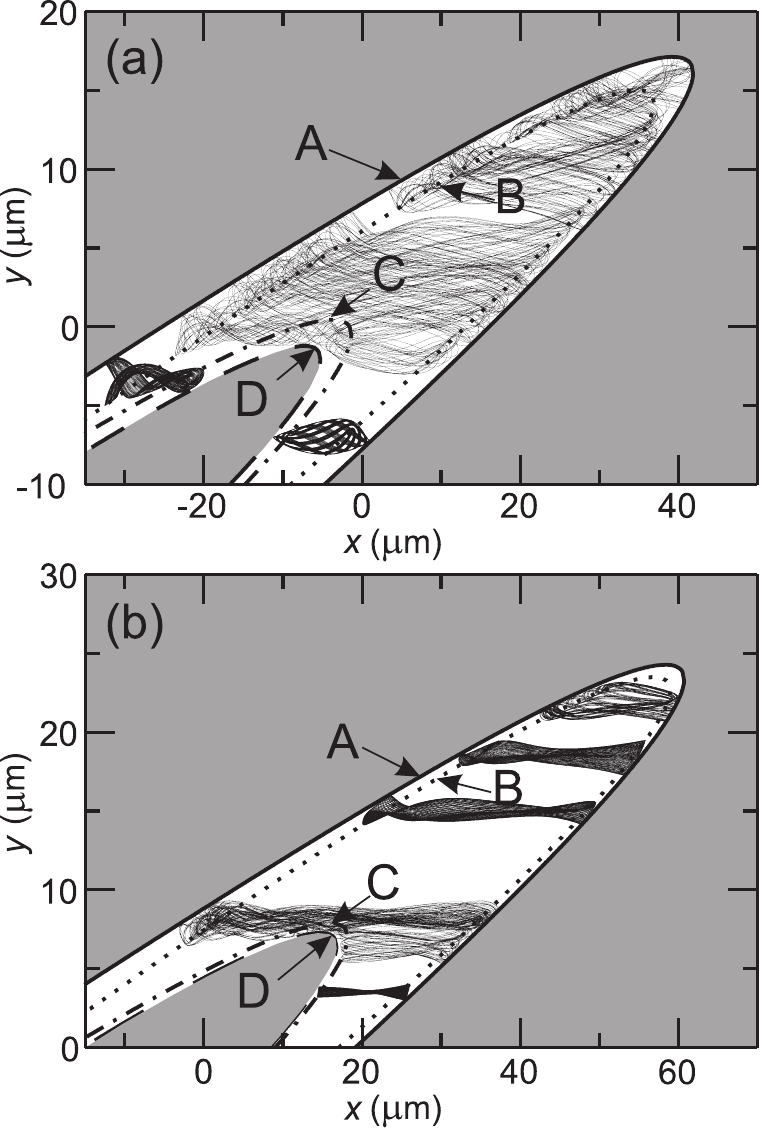}
  }
	\caption{A selection of rays in the crystal with $\omega$ = (a) $6.15\times10^{14}$ rad s$^{-1}$, (b) $6.87\times10^{14}$ rad s$^{-1}$. Also shown are the loci of points in the crystal where $\omega = \omega_{\mathrm{loc}}$ at points A (solid curve), B (dotted curve), C (dot-dash curve) and D (dashed curve) in the reduced Brillouin zone shown in  Fig. \ref{fig2}. In the gray regions there are no propagating solutions at the given $\omega$.}
	\label{fig8}
\end{figure}

The dashed curve in Fig. \ref{fig5} (a,b) is also the boundary between regions where waves can (white) and cannot (gray) propagate. Note that for $y \lesssim -2$ $\mu$m, this separates the allowed region into two segments that occupy distinct ranges of $x$ values. As noted above, when rays approach the dashed curve in Fig. \ref{fig5} they Bragg reflect. Since this boundary is so strongly curved, it makes the ray path highly sensitive to the initial conditions, thereby driving paths that reflect from it chaotic. 

These ray trajectories have some similarities with the classical orbits of particles in a Sinai billiard comprising a rectangular outer wall, which encloses a circular scattering barrier. Particles within the billiard move between these two impenetrable walls. Orbits that interact with the circular barrier are chaotic because reflections from that barrier couple the motion parallel and perpendicular to the outer wall. Similarly, in the photonic crystal considered here, chaotic ray orbits are generated by reflections near the dashed boundary labelled D in Fig. \ref{fig5}.

By contrast, the trajectory of the right-hand path in Fig. \ref{fig5} (a) and enlarged in part (c), is driven chaotic by reflection from the concave (right-hand) end of the white (classically-allowed) region. In Bunimovich stadia, comprising straight parallel edges with semicircular walls at either end, chaos occurs due to reflections from the curved ends, which mix motion parallel and perpendicular to the side walls. In a similar way, ray paths confined near the concave right-hand end of the white allowed region in Fig. \ref{fig5} are also driven chaotic. 

Rays that do not encounter either the concave end of curve A or the convex region around barrier D remain stable, and hence form a dynamical barrier that separates the two chaotic paths in Fig. \ref{fig5}.

Figure \ref{fig6} is a Poincar\'e section in which a point is plotted each time a ray turns from traveling in the negative $x$-direction to the positive $x$-direction. Points corresponding to three different ray paths are shown. The crosses are generated by the upper of the two ray paths in Fig. \ref{fig5} (a), whilst the circles correspond to the lower ray path. The broad spread of these two sets of points in the Poincar\'e section confirms that each ray is chaotic. The triangular points in Fig. \ref{fig6} are generated by a stable ray path (not shown) forming the dynamical barrier that separates the two chaotic orbits in Fig. \ref{fig5}(a). Since this ray is stable, it generates points on a smooth continuous curve in the Poincar\'e section.

To fully understand the origin of the dynamical barrier, it is useful to consider in more detail the bandstructure changes, throughout the crystal, produced by the variation of $l_{x}$ in the $x$ direction, and $m$ with $y'$. 

Figure \ref{fig7} shows dispersion surfaces, $\omega_\mathrm{loc}(\mu_{x},\mu_{y})$, and iso-frequency contours at, (a-c), three different spatial points in the crystal. It illustrates how variations in $l_{x}$ and $m$ both change the bandstructure. The effect of changing $l_{x}$ can be seen by comparing Figs. \ref{fig7}(a) and (b). In (a) $x=0$ $\mu$m, where $l_{x} = l_{0}$, and in (b) $x=50$ $\mu$m, where $l_{x} =$ 0.607 $l_{0}$. In both cases, $y'=0$, and $m=2$  $\mu$m. Reducing $l_{x}$, i.e. moving from (a) to (b), shifts the band to a higher range of frequencies. In addition, reducing $l_{x}$ makes $\omega_\mathrm{loc}$ far less sensitive to changes in $\mu_{y}$, as revealed by comparing Fig. \ref{fig7}(b) with Fig. \ref{fig7}(a).

This reduced variation of $\omega_\mathrm{loc}$ with $\mu_{y}$ means that far more iso-frequency contours form open curves in Fig. \ref{fig7}(b) than in Fig. \ref{fig7}(a), and that the curvature of these contours decreases. The reduction of the curvature of the contours restricts the range of spatial directions in which a ray may propagate. For example, in Fig. \ref{fig7}(b), most rays propagate in directions almost parallel to the $x$ axis.

The effect of changing $m$ on $\omega_\mathrm{loc}(\mu_{x},\mu_{y})$ may be seen by comparing Figs. \ref{fig7}(a) and (c), for which $m=2$ $\mu$m and $m=1$ $\mu$m respectively with fixed $l_{x} = 1$ $\mu$m. As $m$ decreases, the band shifts to higher values of $\omega_\mathrm{loc}$ and broadens so increasing the range of frequencies for which closed contours are found about points A and D. 

We now use these changes in the bandstructure to explain more fully the ray dynamics. Figure \ref{fig8} shows rays in the crystal calculated for  $\omega =$ (a) $6.15\times10^{14}$ rad s$^{-1}$ and (b) $6.87\times10^{14}$ rad s$^{-1}$. The solid, dotted, dot-dashed and dashed curves show, respectively, the loci where $\omega_{\mathrm{loc}}$ coincides with points A, B, C, and D in the Brillouin zone (Figs. \ref{fig2} and \ref{fig7}).

The first point to note is that the (white) region in Fig. \ref{fig8}(b) in which the rays may propagate is located at higher $x$ values than in Fig. \ref{fig8}(a). This is because the band spans higher frequencies when $l_{x}$ is smaller (see Fig. \ref{fig7}), i.e. when $x$ is larger, and so at higher $\omega$ the rays propagate further along the $x$ axis.

Comparing Fig. \ref{fig8}(a) and (b) reveals that at higher $\omega$ [Fig. \ref{fig8}(b)] each ray is more confined in the $y$-direction. This is a consequence of the rays propagating at angles that are closer to the $x$ axis at higher $\omega$; an the effect that can be understood by comparing Figs. \ref{fig7}(a) and (b). Hence, the higher the value of $\omega$, the greater the restriction on the propagation angles, and the narrower the ray path along the $y$ direction. At higher $\omega$, the regions of the crystal in which rays can travel in any direction, i.e. the regions between curves A and B, and curves C and D, are reduced. This corresponds to the reduced variation of $\omega_\mathrm{loc}$ with $\mu_{y}$ at higher $\omega$, as shown in Fig. \ref{fig7}(b).

If $\omega$ is sufficiently low, for this structure when $\omega \lesssim 5.91\times10^{14}$ rad s$^{-1}$, rays in the crystal encounter both of the regions (i.e the concave and convex barriers in Fig. \ref{fig5}) that generate chaos. At higher $\omega$, the spread of each ray orbit in the $y$ direction is reduced, and a ray that encounters one of the regions where it is driven chaotic does not reach the other. When $\omega \gtrsim 6.12\times10^{14}$ rad s$^{-1}$, a region forms in the crystal where rays will encounter neither of the strongly curved barriers. These rays are stable and collectively form the dynamical barrier.

As $\omega$ rises above $\approx 6.6\times10^{14}$ rad s$^{-1}$, the two distinct regions of the crystal that induce chaotic ray paths shrink, and become increasingly localized near the concave and convex barriers. This increases the separation of the two sets of chaotic orbits in real space. If $\omega$ is increased further, the right-hand chaotic region disappears completely, as shown in Fig. \ref{fig8}(b), meaning that the dynamical barrier also vanishes. An examination of rays near the right-hand concave barrier [within the right-hand box in Fig. \ref{fig5}(a)] reveals that, due to the shifting bandstructure throughout the crystal and the effective force that this exerts on the rays, such rays are more complex than those in Bunimovich stadia. Consequently, the exact reason for the suppression of chaos in this region, which occurs at high $\omega$, is not yet fully understood. However, as the ray paths become increasingly restricted to a small region around the center of the concave barrier, chaos disappears because the barrier acts on the orbits as though it is locally flat. By contrast, the chaotic orbits created by the left-hand convex barrier in Fig. \ref{fig5} persist for even higher $\omega$.

\section{Proposed experimental realization of the modulated photonic crystal and associated ray dynamics}
\label{sec:experiment}

Microwave analogues offer major advantages for investigating the propagation of electromagnetic waves through dielectric media and the relation between the waves and underlying ray dynamics \cite{Stoeckmann}. Due to their large scale, photonic crystals for microwaves can be made very accurately with imperfections far smaller than the wavelength of the microwaves. Such imperfections therefore have a negligible effect on the wave transmission processes. For example, modern computer-controlled milling machines can produce holes in a Teflon dielectric block with a precision of $\pm 0.05$ mm: orders of magnitude smaller than the wavelength of microwaves. In addition, the electric field profile of the electromagnetic modes can be directly measured in microwave systems \cite{Stoeckmann}, thus enabling study of, for example, the link between modal form and the transmission coefficient of the system. 

Recent experiments on the transmission of microwaves through 1D photonic crystals comprising Teflon sheets have revealed that absorption of the microwaves by the Teflon has little effect on the shape of the measured transmission spectrum, which is in excellent quantitative agreement with the corresponding calculations \cite{StockNJP}. Since absorption does not destroy coherent wave phenomena, in particular band formation, we expect that it will have a similarly small effect on the band transport processes discussed above. Consequently, modulated microwave photonic crystals fabricated by milling cm-scale air holes into a Teflon sheet, or high-refractive-index (3.3) ceramic-filled plastic \cite{ceramic}, offer a promising starting point for the experimental verification and study of the dynamical effects discussed above, and their effect on electromagnetic wave transmission. We note that photonic crystals with a gradient index have been successfully realized experimentally for metamaterials \cite{Mei_2009,Chen_2009,Liu_2009,Cai_2007,Schurig_2006,Smith_2005}, and such systems might also be suitable for experimental studies of band transport phenomena in modulated photonic crystals.

Here, though, we focus on possible microwave realizations of the modulated photonic crystals. To ensure maximal correspondence with the system shown in Fig. \ref{fig5}, and analyzed in detail above, the microwave photonic crystal would have rectangular air holes separated by thin dielectric walls. The spatial variation of the hole and wall widths follows from the requirement that $l_{x} = l_{0}\exp(-\eta{x})$ and $m=(\epsilon_{r}-1)d=m_{0}(1-\rho{y'}^{2})$, as discussed in Section \ref{sec:The structure of the crystal}. Providing the walls are much narrower than the air holes, the local dispersion relation of the frequency bands will be accurately described by Shepherd's analytical model \cite{shep1} (see Section \ref{sec:The structure of the crystal}). We note, however, that exact dispersion relations can be determined for any unit cell geometry by using standard techniques such as expansion over plane waves.     

As in previous experiments \cite{Stoeckmann,Stein_1995}, quasi 2D microwave photonic crystals could be used because the width of the system along $z$ does not significantly affect either the frequency band structure or the resulting ray dynamics. In such structures, the dielectric block is $\approx 1$ cm thick along the $z$ direction and enclosed by copper plates, which form a microwave cavity. To study transport through the 2D photonic crystal, continuous microwaves would be injected through a wire antenna, of diameter $\approx$ 0.1 mm, which passes through the top plate at a fixed position, $\bm{r}$. A vector network analyzer would be used to measure the spectrum $T(\omega, \bm{r}, \bm{r'})$ for transmission to a receiving wire antenna at a variable position $\bm{r'}$ in the bottom plate for $\omega$ in the range $30 \times 10^{12} \lesssim \omega \lesssim 250 \times 10^{12}$ rad s$^{-1}$ \cite{Stoeckmann,Stein_1995}. 

A key advantage of using a vector network analyzer is that for each $\omega$ value, the electric field profile, $E(x, y)$, can be mapped experimentally by using a moveable bottom plate to scan the receiving wire across the structure and analyzing how $T(\omega, \bm{r}, \bm{r'})$ changes with $\bm{r'}$ \cite{Stoeckmann}. Since both the modulus and sign of $E(x, y)$ can be determined, continuous waves can study pulse propagation via a simple Fourier transform of $T(\omega, \bm{r}, \bm{r'})$ to the time domain \cite{Stein_1995}. This Fourier technique, which is simpler than generating microwave pulses \cite{Stoeckmann}, could be used to investigate the effects of dynamical barriers on pulse propagation by measuring how $T(\omega, \bm{r}, \bm{r'})$ varies with the positions of the two antennae, with $\omega$, and with $\theta$. 

For $\theta$ values that generate a dynamical barrier, which separates chaotic trajectories into distinct regions of phase space (Figs. \ref{fig5} and \ref{fig6}), the dynamical barrier will reduce the range of ray trajectories that directly link the two antennae, thereby producing a measurable decrease in transmission between them. As shown in Fig. \ref{fig8}, the location, and even the existence, of the dynamical barrier varies rapidly with $\omega$. Consequently, we expect that $T(\omega, \bm{r}, \bm{r'})$ will also demonstrate high sensitivity to changes in $\omega$. Providing the parameters of the photonic crystal are accurately known, we expect good quantitative agreement between experiment and theory, which will help to elucidate the effect of the dynamical barriers on ray and wave transmission. 

The aim of the proposed microwave experiments is to investigate the feasibility of using dynamical barriers to control electromagnetic wave propagation through spatially-modulated photonic crystals and so provide proof-of-principle demonstration of this concept. We emphasize that dynamical barriers occur in many systems with a mixed stable-chaotic classical phase. In the electromagnetic system that we consider here, their existence requires spatial modulation of the bandstructure, which couples motion in the $x$ and $y$ directions, but does not depend on the precise details of the lattice structure used to produce that modulation: for example the shape of the unit cell. Consequently, the dynamics that we consider have the potential for scaling to optical wavelengths, where the air holes are usually circular rather than rectangular. A rectangular unit cell speeds up the ray calculations, making it easier to explore the large parameter space available, but is not essential to generate dynamical barriers, which should occur in a range of modulated photonic crystal structures.

\section{Summary}
\label{sec:Summary}

We have shown that spatial modulation of the lattice parameters and, hence, the local bandstructure of a photonic crystal affects the stability, form, and location of Hamiltonian ray paths within it. The formation of convex or concave barriers at the edges of the classically-allowed region drives orbits that interact with those barriers chaotic. As $\omega$ increases, changes in the dispersion relation for the second allowed band make the ray paths increasingly localized in the $y$ direction. In turn, this confines the chaotic ray trajectories to smaller areas around the convex and concave barriers at the edges of the classically-allowed region. The location and extent of the chaotic trajectories depend strongly on the value of $\omega$. For certain $\omega$, only a single region of chaos exists. However, as $\omega$ increases, this splits into two distinct regions of chaotic ray paths, located near either the convex or the concave barrier. These chaotic regions are separated by a dynamical barrier comprising stable ray orbits. When $\omega$ is just large enough for the dynamical barrier to form, this barrier is leaky because, in some regions of the crystal, chaotic orbits that are isolated in real space overlap in phase space. But as $\omega$ increases, the chaotic regions separate fully in both real and phase space. In this regime, the chaotic orbits cannot penetrate the dynamical barrier and are thus unable to propagate through the crystal. As $\omega$ increases further, the chaotic rays become so localized that those near the concave barrier disappear completely, thus removing the dynamical barrier. 

We conclude that rays in slowly-modulated photonic crystals exhibit unique and rich dynamics, which are highly sensitive to both the spatial structure of the crystal and the value of $\omega$. Dynamical barriers formed by the stable rays provide a mechanism for controlling light transmission in certain frequency ranges. The exponential separation rate of two chaotic rays with slightly different initial conditions also strongly affects the propagation of the electromagnetic radiation. This mechanism could be exploited in optical switches or routers and may also enhance angular dispersion due to the superprism effect in photonic crystal structures \cite{x1}, which has potential for applications in wavelength division multiplexing devices. We hope that our results will stimulate the further work required to determine how \emph{ray} chaos affects the transmission of electromagnetic \emph{waves} through modulated photonic crystals, and to test the validity of Hamiltonian optics in this complex regime of nonlinear dynamics. 


\end{document}